\documentclass[10pt, conference, compsocconf]{IEEEtran}
\usepackage{cite}
\usepackage{amsmath,amssymb,amsfonts}
\usepackage{algorithmic}
\usepackage{graphicx}
\usepackage{textcomp}
\usepackage{xcolor}
\usepackage[colorlinks=true,linkcolor=blue]{hyperref}
\hypersetup{
  colorlinks,
  citecolor=blue,
  linkcolor=Red,
  urlcolor=blue}
\begin{document}

\title{On Continuous Integration / Continuous Delivery for Automated Deployment of Machine Learning Models using MLOps}
\author{\IEEEauthorblockN{Satvik Garg\textsuperscript{1}, Pradyumn Pundir\textsuperscript{1}, Geetanjali Rathee\textsuperscript{2}, P.K. Gupta\textsuperscript{1}, Somya Garg\textsuperscript{3}, Saransh Ahlawat\textsuperscript{3}}
\IEEEauthorblockA{\textsuperscript{1}Jaypee University of Information Technology, Solan, India\\
\textsuperscript{2}Netaji Subhas University of Technology, New Delhi, India\\
\textsuperscript{3}Deloitte LLC, New York, NY, USA\\
Email: {\{satvikgarg27, pundirpradyumn25, geetanjali.rathee123\}@gmail.com}, pkgupta@ieee.org,\\ \{somgarg, saahlawat\}@deloitte.com}}

\maketitle
\begin{abstract}
Model deployment in machine learning has emerged as an intriguing field of research in recent years. It is comparable to the procedure defined for conventional software development. Continuous Integration and Continuous Delivery (CI/CD) have been shown to smooth down software advancement and speed up businesses when used in conjunction with development and operations (DevOps). Using CI/CD pipelines in an application that includes Machine Learning Operations (MLOps) components, on the other hand, has difficult difficulties, and pioneers in the area solve them by using unique tools, which is typically provided by cloud providers. This research provides a more in-depth look at the machine learning lifecycle and the key distinctions between DevOps and MLOps. In the MLOps approach, we discuss tools and approaches for executing the CI/CD pipeline of machine learning frameworks. Following that, we take a deep look into push and pull-based deployments in Github Operations (GitOps). Open exploration issues are also identified and added, which may guide future study.
\end{abstract}


\section{Introduction}
Artificial Intelligence (AI) is a field of study that has been on the ascent to develop applications fit for procuring information and providing models dependent on past knowledge \cite{langley1995applications}. 
For real-world applications, the data is so vast and made available in the production system only, which brings about the management of various tools, including libraries, and dependencies that can complicate model development \cite{lwakatare2020large}. Therefore, it has been valuable to incorporate the concepts of continuous training and continuous development (CI/CD) for automated deployment of AI models \cite{garg2019automated}. 

In real-world systems, machine learning (ML) models must be flexible according to changing input data. Once the model has been deployed to production, the performance of the model degrades due to frequent changes to the data \cite{wu2020deltagrad}. Therefore, model monitoring is essential and responsible for adequately monitoring the pipeline that can trigger, retrain and ensure models are working as expected. However, tracking the performance of a model comes with several challenges throughout the machine learning lifecycle \cite{sculley2015hidden}. All of the these challenges can be solved using MLOps \cite{alla2021mlops}. What is MLOps? 
In simple terms, MLOps is DevOps for machine learning. It enables developers
to collaborate and increase the pace at which AI models can be developed, deployed, scaled, monitored, and retrained. 

The idea behind presenting such a paper is to put together an investigation that gathers information on DevOps methodology for machine learning applications. On looking through research papers and interacting with experts in the field of CI/CD, we found a knowledge gap among machine learning developers in building automation pipelines 
\cite{wan2019does} \cite{wolf2020sensemaking}. Our most significant commitment to this paper is not a methodology but an investigation of approaches. It covers using existing advances, for example, containerization tools such as docker \cite{Empoweri53:online}, orchestration tools like Kubernetes \cite{sayfan2017mastering}, and MLOps platforms such as Amazon SageMaker \cite{AmazonSa87:online}, KubeFlow \cite{Introduc29:online}, and MLFlow \cite{zaharia2018accelerating} that are demanding and easy to implement. Subsequently, we provide a bird's eye view on the Github operations \cite{GitOpsGi48:online} related to model deployment to help guide new researchers. 

The rest of the sections are organized as follows: 
Section 2 introduces the concepts of DevOps, including concepts of CI/CD, containerization, and orchestration. Section 3 involves MLOps levels and discusses the key differences between MLOps and DevOps. Section 4 provides information on Github operations for testing push and pull-based deployments. Section 5 explores open research challenges, and Section 6 discusses the related works. Section 7 concludes the paper.

\section{Development and Operations}
DevOps or development and operations can be termed as the practice used by an organization while developing software applications\cite{garg2019automated}. 
There are mainly two concepts in DevOps:
\begin{itemize}
    \item Continuous Integration (CI): 
    CI helps in time management and enables an organization to have short and frequent release cycles for improving software quality and increase overall team productivity.
    \item Continuous Deployment (CD): It helps automatically deploy software in production \cite{singh2019comparison}. The main difference between CI and CD is mainly about ensuring an application availability in production ready condition after performing quality checks. 
\end{itemize}





Machine learning pipelines require components to be reusable, composable and potentially shareable across the pipeline. These components should ideally be containerized to decouple execution environment for custom code runtime and make the code reproducible between developers
\cite{turnbull2014docker}.
Docker \cite{Empoweri53:online} has been widely used for containerization. 
As shown in Figure 1, the docker platform provides the ability to package and run applications in a small isolated environment by using a client-server architecture, where the client communicate to docker daemon, for processes such as building, running, and distributing docker containers and performs essential functions. 
However, for large scale applications, thousands of containers need to be deployed which can be challenging to manage. Orchestration solves the problem of docker adoption by allowing developers to leverage the concepts of container automation, deployment, and networking \cite{rufino2017orchestration}. In most of the cases, the docker containers are organized by kubernetes \cite{sayfan2017mastering}.
\begin{figure}[htbp]
\centerline{\includegraphics[width=8cm, height=2.8cm]{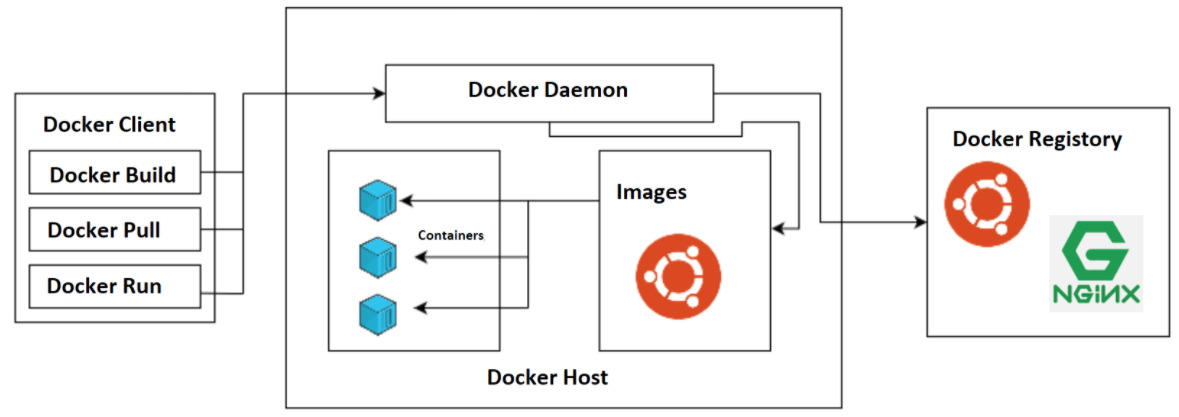}}
\caption{Docker Containerization.}
\end{figure}


\section{Machine Learning Operations}


The task of deploying AI models to production requires processes after the business model has been determined to ascertain the success rate. These processes provide ML models for production and can be accomplished manually or with the aid of automated workflows. As a result, three different levels of MLOps are defined namely, MLOps Level 0, MLOps Level 1, MLOps Level 2 \cite{MLOpsCon49:online}. 

\subsection{MLOps Level 0}
The fundamental level of maturity, or MLOps level 0, refers to simple workflows that are manually script-driven at every stage of the machine learning lifecycle. 
This level of MLOps suffers from sparse release iterations because it expects the generated model to not change very often. The concepts of CI/CD are not needed as there is no automation, resulting in lack of active performance monitoring. 
When models are repeatedly changed or trained, the manual-driven process may persist, but in practice, models are regularly changed or trained, demanding regular iterations.



\subsection{MLOps Level 1}

The steps of machine learning experiments are orchestrated at MLOps level 1, to automate the ML pipeline solely to undertake continuous model training. This enables it to provide model prediction services continually and automates retraining models in production using new data. 
As a result, model deployment setups and continuous model delivery are automated, allowing to leverage trained and validated models as a prediction service for making the online predictions. However, in order to test new ideas and quickly release new implementations of ML components, the need for CI/CD solutions to automate the creation, testing and deployment of ML pipelines is critical.


\subsection{MLOps Level 2}
Level 2 of MLOps includes automation of a CI/CD pipeline for more rapid and reliable updating of pipelines in production, which requires a more robust automated system. As shown in Figure 2, 
a total of six CI/CD processes has been incorporated that includes development and experimentation, where staged experiment phases can be carried out iteratively for training new algorithms and models \cite{MLOpsCon49:online}. It results in the output of pipeline steps' source code, subsequently pushed into the source repository. Next step involves continuous integration (CI), where the source code is built and undergoes various run tests, resulting in package executables and artifacts being deployed later. Finally, in continuous delivery (CD), the artifacts created in CI phases are deployed to the target environment, resulting in the implementation of updated AI models through the generated pipeline. This pipeline is automated in production and runs according to a schedule or a trigger. The generated trained models are then uploaded to the model registry, and this stage is also known as automatic triggering. It also includes CD, which serves prediction as a service. So, once a model prediction service has been installed, the statistics on model performance based on real-time data are collected. The result of this stage is a trigger to execute pipelines and start a new trial cycle.

\begin{figure}[htbp]
\centerline{\includegraphics[width=6cm, height=7cm]{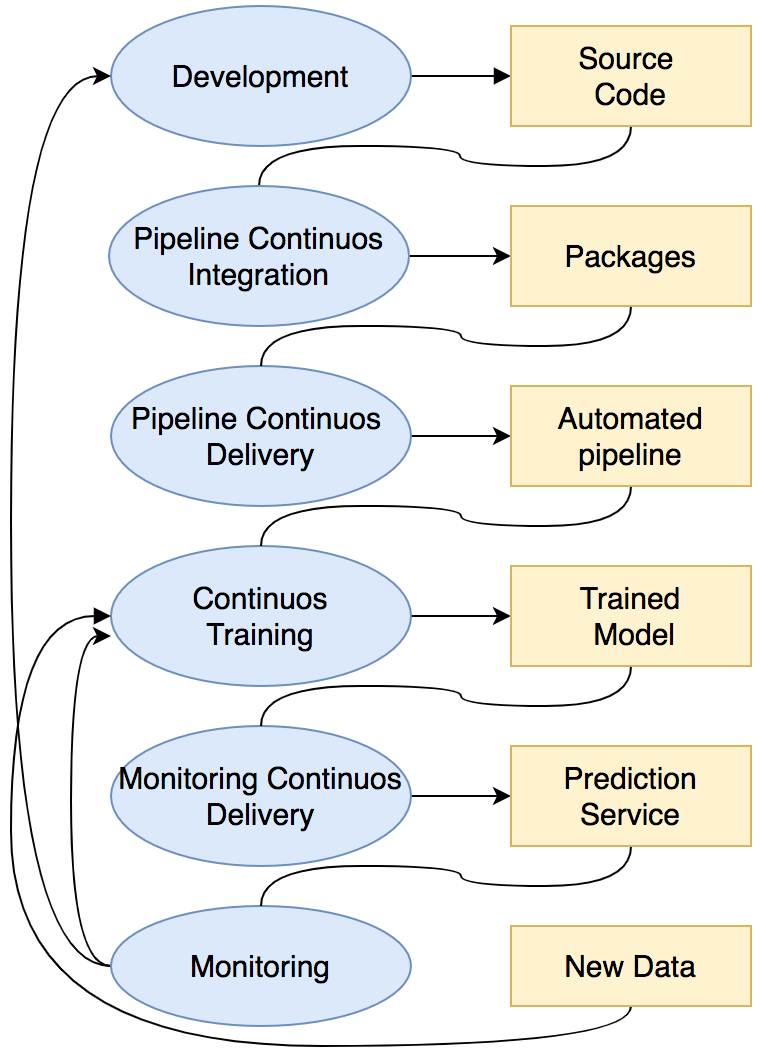}}
\caption{MLOPS Automated CI/CD Pipeline \cite{MLOpsCon49:online}}
\end{figure}

\subsection{MLOps vs DevOps}
MLOps is DevOps to ML, but only to a certain extent, and there are key differences that needs to be addressed by the MLOps platforms \cite{karamitsos2020applying}.
\begin{itemize}
    \item In a typical software application, if an error needs to be fixed, the task is only to edit the code, test, and deploy it. Nonetheless, model training in ML is an experimental field that requires various efforts on alternative data parameters, algorithms, and feature engineering techniques. 
    \item Integration and unit testing are of central significance to guarantee the right execution of software ecosystems. In any case, the different testing procedures applied in software systems are not usually implemented in ML-based systems.  
    \item When a model is put into production, it begins to make predictions based on the knowledge gained from real-time data. As business goals adjust, this knowledge is modified which leads to model degradation. Different types of monitoring, such as covariate-shift and pre-shift \cite{Detectin39:online} needs to be implemented, unlike traditional software applications, which in most cases involves monitoring latency. 
\end{itemize}

\subsection{MLOps Platform}
Recent years have seen a plethora of MLOps platforms
to provide standards for deploying enterprise level AI applications 
\cite{Introduc29:online, AmazonSa87:online, zaharia2018accelerating}. These platforms are open-source projects that contains curated set of compatible tools and frameworks for machine learning to ease the process of automated development. Kubeflow \cite{Introduc29:online} introduces the concept of the namespaces that enables multi-user support to simplify collaboration and access management.
In addition, it includes a built-in Jupiter notebook environment in the clusters which makes it simple to perform jobs with dynamically scaled resources. Katib, which can be utilized for automated hyperparameter adjustment, is another crucial component in kubeflow, 
responsible for determining and visualizing the best configuration for the model before it goes into production.

Amazon SageMaker \cite{AmazonSa87:online} 
allows connecting and loading data from sources such as Amazon S3 bucket, Amazon Athena, and Amazon Redshift. 
that leverages data manager, which gives the function of combining and merging original data characteristics within a few seconds. 
It is essential to make sure that the data and features are well balanced and defined. For this, the SageMaker Clarify can be used to verify each prediction. Finally, the Sage Maker debugger can remove them from the model to identify the source of errors.

In MLflow \cite{zaharia2018accelerating}, the MLflow tracking helps to recover and query experimental data and results. Tools like ML projects and MLflow models allow implementing ML models in the environment. A model registration feature helps to store and manage models in a central repository. In addition, it has built-in integration with TensorFlow, PyTorch, Keras, python, Kubernetes, docker, etc. Many organizations such as Databricks, Toyota, Accenture, Facebook, and Microsoft are employing and contributing to MLflow, which easily provides community support for resolving bugs.


\section{Github Operations}

GitOps or Github Operations has been considered as a continuous deployment method for cloud-native applications \cite{limoncelli2018gitops}. It can also be termed a delivery tool that uses a pull-based model, in contrast, employing a push-based model. The core idea of GitOps is to leverage the Git repository which contains the details of the infrastructure required in a production environment. It has an automated process to handle necessary steps required for deployment. For example, it can deploy a new application or update an existing application by simply updating the repository. It accesses a repository or an image registry, and users can give developers direct access to the environment. 
Tools such as ArgoCd, Flux and Fleet are developed to maintain the GitOps workflow. 

GitOps supports the management of multiple pipelines by configuring different build pipelines to update the environment repository and helps to configure an operator or deployment pipeline to react to changes in a branch \cite{GitOpsGi38:online}. The deployment approach for GitOps can be implemented in two ways: push-based and pull-based deployments. The difference between these two deployment types is how the deployment environment matches the desired infrastructure. Pull-based approaches should be chosen whenever possible because they are deemed more secure and better practices for implementing GitOps.

\section{Open Research Challenges}
Over time organizations have noticed the significance of MLOps in executing a constructive production pipeline. However, MLOps has still in its infancy, and most firms are still working on the best way to adapt it for their specific projects \cite{lenarduzzi2021software}. 
\begin{itemize}
\item An issue that develops in the early stages is about monitoring the effectiveness of AI models and a poor understanding of solution needs can hurt solution performance. The changes in information are more frequent that can disrupt the entire model even though the collected data is free from errors. 
\item The developers are still more inclined towards manually testing and deploying AI models in production as they are not assertive on automatically delivering each change.
\item To comprehend and port the code from one library to the next is regularly nontrivial. The ascent of numerous libraries for AI development and the distinctive foundation knowledge of the specialists produced challenges for coding. Libraries, for example, Tensorflow, Keras, and Pytorch have alternate spines which makes a similar application appear to be unique from each other.
\item In most of the cases, the development and production teams are out of sync and communicate only at the end of solution design. An approval is required in case someone on the team changes a requirement that has to be reflected on the production server making the whole process slow.
\end{itemize}

\section{Related Works}

The literature \cite{renggli2019continuous} describes an optimized approach to implement MLOps using various platforms and tools to automate the execution of AI models. 
The computer vision-based model has been adapted into manufacturing measures by incorporating the MLOps-based application lifecycle \cite{lim2019mlop}. 
Also, the sudy paper in \cite{karamitsos2020applying}, suggests techniques for applying DevOps and CI/CD in AI applications.
Some of the recent works features the issue of quality in AI software \cite{murphy2006framework, zhang2020machine}. 
It is also important to focus product quality assurance on the ML framework \cite{chen2015metamorphic}. 
The authors in \cite{wang2020better} investigates practices for using Jupyter notebooks and suggests to identify new methods to analyze these notebooks. The proposed literature \cite{QualityA28:online} aims to repeatedly test the post-deployment quality of AI models on real-time data considering a black-box approach. 

\section{Conclusion}
As part of this exploratory research, we gathered information about the current MLOps, and the key motivators behind having operational processes that can facilitate clear and efficient workflow for delivering machine learning solutions. We inherited ideas from traditional software development, specifically the CI/CD techniques and presented use-cases for when those practices could be valuable in the machine learning development lifecycle. We proposed 3 different levels of MLOps that vary in efficiency and the amount of resources required to set them up, because the understanding is that not every ML solution would require similar amount of provisioning. The level 3 of MLOps is the most sophisticated solution proposed and we recommend using it in any enterprise machine learning solution. This level of granularity would achieve the most efficient and automated development workflow as it leverages the industry standard tools and techniques that have successfully delivered enterprise level software solutions in the past decade. In future work, we plan to produce a demo using a proof of concept applications to compare all three approaches for different size of applications.

\bibliography{sample.bib}

\begin{thebibliography}{10}
\providecommand{\url}[1]{#1}
\csname url@samestyle\endcsname
\providecommand{\newblock}{\relax}
\providecommand{\bibinfo}[2]{#2}
\providecommand{\BIBentrySTDinterwordspacing}{\spaceskip=0pt\relax}
\providecommand{\BIBentryALTinterwordstretchfactor}{4}
\providecommand{\BIBentryALTinterwordspacing}{\spaceskip=\fontdimen2\font plus
\BIBentryALTinterwordstretchfactor\fontdimen3\font minus
  \fontdimen4\font\relax}
\providecommand{\BIBforeignlanguage}[2]{{%
\expandafter\ifx\csname l@#1\endcsname\relax
\typeout{** WARNING: IEEEtran.bst: No hyphenation pattern has been}%
\typeout{** loaded for the language `#1'. Using the pattern for}%
\typeout{** the default language instead.}%
\else
\language=\csname l@#1\endcsname
\fi
#2}}
\providecommand{\BIBdecl}{\relax}
\BIBdecl

\bibitem{langley1995applications}
P.~Langley and H.~A. Simon, ``Applications of machine learning and rule
  induction,'' \emph{Communications of the ACM}, vol.~38, no.~11, pp. 54--64,
  1995.

\bibitem{lwakatare2020large}
L.~E. Lwakatare, A.~Raj, I.~Crnkovic, J.~Bosch, and H.~H. Olsson, ``Large-scale
  machine learning systems in real-world industrial settings: A review of
  challenges and solutions,'' \emph{Information and Software Technology}, vol.
  127, p. 106368, 2020.

\bibitem{garg2019automated}
S.~Garg and S.~Garg, ``Automated cloud infrastructure, continuous integration
  and continuous delivery using docker with robust container security,'' in
  \emph{2019 IEEE Conference on Multimedia Information Processing and Retrieval
  (MIPR)}.\hskip 1em plus 0.5em minus 0.4em\relax IEEE, 2019, pp. 467--470.

\bibitem{wu2020deltagrad}
Y.~Wu, E.~Dobriban, and S.~Davidson, ``Deltagrad: Rapid retraining of machine
  learning models,'' in \emph{International Conference on Machine
  Learning}.\hskip 1em plus 0.5em minus 0.4em\relax PMLR, 2020, pp.
  10\,355--10\,366.

\bibitem{sculley2015hidden}
D.~Sculley, G.~Holt, D.~Golovin, E.~Davydov, T.~Phillips, D.~Ebner,
  V.~Chaudhary, M.~Young, J.-F. Crespo, and D.~Dennison, ``Hidden technical
  debt in machine learning systems,'' \emph{Advances in neural information
  processing systems}, vol.~28, pp. 2503--2511, 2015.

\bibitem{alla2021mlops}
S.~Alla and S.~K. Adari, ``What is mlops?'' in \emph{Beginning MLOps with
  MLFlow}.\hskip 1em plus 0.5em minus 0.4em\relax Springer, 2021, pp. 79--124.

\bibitem{wan2019does}
Z.~Wan, X.~Xia, D.~Lo, and G.~C. Murphy, ``How does machine learning change
  software development practices?'' \emph{IEEE Transactions on Software
  Engineering}, 2019.

\bibitem{wolf2020sensemaking}
C.~T. Wolf and D.~Paine, ``Sensemaking practices in the everyday work of ai/ml
  software engineering,'' in \emph{Proceedings of the IEEE/ACM 42nd
  International Conference on Software Engineering Workshops}, 2020, pp.
  86--92.

\bibitem{Empoweri53:online}
``Empowering app development for developers | docker,''
  \url{https://www.docker.com/}, (Accessed on 09/13/2021).

\bibitem{sayfan2017mastering}
G.~Sayfan, \emph{Mastering kubernetes}.\hskip 1em plus 0.5em minus 0.4em\relax
  Packt Publishing Ltd, 2017.

\bibitem{AmazonSa87:online}
``Amazon sagemaker – machine learning – amazon web services,''
  \url{https://aws.amazon.com/sagemaker/}, (Accessed on 09/13/2021).

\bibitem{Introduc29:online}
``Introduction to kubeflow | kubeflow,''
  \url{https://www.kubeflow.org/docs/about/kubeflow/}, (Accessed on
  09/02/2021).

\bibitem{zaharia2018accelerating}
M.~Zaharia, A.~Chen, A.~Davidson, A.~Ghodsi, S.~A. Hong, A.~Konwinski,
  S.~Murching, T.~Nykodym, P.~Ogilvie, M.~Parkhe \emph{et~al.}, ``Accelerating
  the machine learning lifecycle with mlflow.'' \emph{IEEE Data Eng. Bull.},
  vol.~41, no.~4, pp. 39--45, 2018.

\bibitem{GitOpsGi48:online}
``Gitops | gitops is continuous deployment for cloud native applications,''
  \url{https://www.gitops.tech/}, (Accessed on 09/02/2021).

\bibitem{singh2019comparison}
C.~Singh, N.~S. Gaba, M.~Kaur, and B.~Kaur, ``Comparison of different ci/cd
  tools integrated with cloud platform,'' in \emph{2019 9th International
  Conference on Cloud Computing, Data Science \& Engineering
  (Confluence)}.\hskip 1em plus 0.5em minus 0.4em\relax IEEE, 2019, pp. 7--12.

\bibitem{turnbull2014docker}
J.~Turnbull, \emph{The Docker Book: Containerization is the new
  virtualization}.\hskip 1em plus 0.5em minus 0.4em\relax James Turnbull, 2014.

\bibitem{rufino2017orchestration}
J.~Rufino, M.~Alam, J.~Ferreira, A.~Rehman, and K.~F. Tsang, ``Orchestration of
  containerized microservices for iiot using docker,'' in \emph{2017 IEEE
  International Conference on Industrial Technology (ICIT)}.\hskip 1em plus
  0.5em minus 0.4em\relax IEEE, 2017, pp. 1532--1536.

\bibitem{MLOpsCon49:online}
``Mlops: Continuous delivery and automation pipelines in machine learning,''
  \url{https://cloud.google.com/architecture/mlops-continuous-delivery-and-automation-pipelines-in-machine-learning},
  (Accessed on 08/13/2021).

\bibitem{karamitsos2020applying}
I.~Karamitsos, S.~Albarhami, and C.~Apostolopoulos, ``Applying devops practices
  of continuous automation for machine learning,'' \emph{Information}, vol.~11,
  no.~7, p. 363, 2020.

\bibitem{Detectin39:online}
``Detecting data drift with mlops | 10clouds,''
  \url{https://10clouds.com/blog/detecting-data-drift-mlops/}, (Accessed on
  09/7/2021).

\bibitem{limoncelli2018gitops}
T.~A. Limoncelli, ``Gitops: a path to more self-service it,''
  \emph{Communications of the ACM}, vol.~61, no.~9, pp. 38--42, 2018.

\bibitem{GitOpsGi38:online}
``Gitops | gitops is continuous deployment for cloud native applications,''
  \url{https://www.gitops.tech/}, (Accessed on 09/15/2021).

\bibitem{lenarduzzi2021software}
V.~Lenarduzzi, F.~Lomio, S.~Moreschini, D.~Taibi, and D.~A. Tamburri,
  ``Software quality for ai: Where we are now?'' in \emph{International
  Conference on Software Quality}.\hskip 1em plus 0.5em minus 0.4em\relax
  Springer, 2021, pp. 43--53.

\bibitem{renggli2019continuous}
C.~Renggli, B.~Karla{\v{s}}, B.~Ding, F.~Liu, K.~Schawinski, W.~Wu, and
  C.~Zhang, ``Continuous integration of machine learning models with ease.
  ml/ci: Towards a rigorous yet practical treatment,'' \emph{arXiv preprint
  arXiv:1903.00278}, 2019.

\bibitem{lim2019mlop}
J.~Lim, H.~Lee, Y.~Won, and H.~Yeon, ``Mlop lifecycle scheme for vision-based
  inspection process in manufacturing,'' in \emph{2019 $\{$USENIX$\}$
  Conference on Operational Machine Learning (OpML 19)}, 2019, pp. 9--11.

\bibitem{murphy2006framework}
C.~Murphy, G.~E. Kaiser, and M.~Arias, ``A framework for quality assurance of
  machine learning applications,'' 2006.

\bibitem{zhang2020machine}
J.~M. Zhang, M.~Harman, L.~Ma, and Y.~Liu, ``Machine learning testing: Survey,
  landscapes and horizons,'' \emph{IEEE Transactions on Software Engineering},
  2020.

\bibitem{chen2015metamorphic}
T.~Y. Chen, ``Metamorphic testing: A simple method for alleviating the test
  oracle problem,'' in \emph{2015 IEEE/ACM 10th International Workshop on
  Automation of Software Test}.\hskip 1em plus 0.5em minus 0.4em\relax IEEE,
  2015, pp. 53--54.

\bibitem{wang2020better}
J.~Wang, L.~Li, and A.~Zeller, ``Better code, better sharing: on the need of
  analyzing jupyter notebooks,'' in \emph{Proceedings of the ACM/IEEE 42nd
  International Conference on Software Engineering: New Ideas and Emerging
  Results}, 2020, pp. 53--56.

\bibitem{QualityA28:online}
``Quality assurance for machine learning models - part 1,''
  https://blog.sasken.com/quality-assurance-for-machine-learning-models-part-1-why-quality-assurance-is-critical-for-machine-learning-models,
  (Accessed on 08/25/2021).

\end{thebibliography}

\end{document}